\documentclass[12pt]{article}

\usepackage{amsmath}
\usepackage{amssymb}

\usepackage{indentfirst}
\usepackage{amssymb}
\usepackage{amsfonts}
\usepackage{amscd}
\usepackage{amsbsy}
\usepackage{amsthm}
\usepackage{latexsym}
\usepackage{graphicx,color} 
\usepackage[dvipsnames]{xcolor}
\usepackage[colorlinks]{hyperref}
\usepackage{latexsym}
\usepackage[dvipsnames]{xcolor}
\usepackage{hyperref}

\hypersetup{linkcolor=blue,citecolor=blue,urlcolor=blue}

\def\nn{\nonumber}       
\def\beq{\begin{eqnarray}}
\def\eeq{\end{eqnarray}}

\def\ln{\,\mbox{ln}\,}

\def\det{\,\mbox{det}\,}
\def\tr{\,\mbox{tr}\,}

\def\Tr{\,\mbox{Tr}\,}

\DeclareMathOperator{\cx}{\square}

\def\al{\alpha}
\def\be{\beta}

\def\ga{\gamma}
\def\de{\delta}

\def\ep{\epsilon}

\def\la{\lambda}
\def\na{\nabla}
\def\pa{\partial}

\def\si{\sigma}

\def\ph{\varphi}

\def\Ga{\Gamma}
\def\De{\Delta}

\usepackage[symbol]{footmisc} 
\usepackage{float} 
\usepackage{cite}  

\usepackage{titlesec}

\titleformat*{\section}{\large\bfseries}
\titleformat*{\subsection}{\normalsize\bfseries}

\begin{document}

\begin{center}
	
{\Large
Conformal anomaly in a vector field model with \\ auxiliary scalar field}
\vskip 6mm
	
\textbf{Samuel W. P. Oliveira}
	\footnote{E-mail address: sw.oliveira55@gmail.com}$^{a}$,
	\quad
\textbf{P\'ublio Rwany B. R. do Vale}\footnote{
E-mail address: publio.vale@gmail.com}$^{b}$
	\\
and \quad
\textbf{Ilya L. Shapiro}\footnote{
		E-mail address: ilyashapiro2003@ufjf.br}$^{c,a}$
\vskip 6mm
	
$a.$ \
PPGCosmo,
	Universidade Federal do Esp\'{\i}rito Santo,
	\\
Vit\'oria, 29075-910, ES, Brazil
\vskip 3mm
$b.$
 International Institute of Physics, Universidade Federal do
 Rio Grande do Norte,
\\	
Natal, 59078-970, RN, Brazil
\vskip 3mm
$c.$ \
Departamento de F\'{\i}sica, \ ICE, \
	Universidade Federal de Juiz de Fora,
\\
Juiz de Fora,  36036-900,  MG,  Brazil
\end{center}
\vskip 10mm

\begin{abstract}
	
\noindent
The conformal anomaly has well-known ambiguities related to the
possible schemes of regularization and renormalization. In case of
dimensional regularization, one of the options is to formulate the
theory as conformal in the dimension $D \neq 4$. For a gauge vector
field this can be done in several ways and one of the options is to
introduce an auxiliary scalar playing the role of a compensator. The
advantage of this approach is that it preserves gauge symmetry and
avoids problems with possible violation of unitarity. We explore the
consequences of introducing such an auxiliary field for the anomaly
and anomaly-induced action. It is shown that  the new scalar degree
of freedom gains an independent dynamics after taking the $4D$ limit.
The remnant scalar, also, demonstrates some interesting properties.
\vskip 3mm

	\vskip 2mm
	
\end{abstract}

\setcounter{footnote}{0}
\renewcommand*{\thefootnote}{\arabic{footnote}}
\newpage

\section{Introduction}
\label{sec 1}

The local conformal symmetry and its quantum violation in the form
of trace anomaly is an object of extensive studies from the epoch of
initial works \cite{CapDuf-74} and \cite{ddi,duff77}. In the
subsequent decades, this area got a lot of importance, in particular
because of the relevant applications in string theory, cosmology and
black hole physics (see, e.g., \cite{duff94} and \cite{PoImpo}).
On the other hand, there are some questions which are difficult to
answer and there is still a possibility to find some new surprising
aspects of the problem. In particular, this concerns the 
ambiguity related to regularization (see, e.g., the discussion in
\cite{birdav,duff94}) which concerns a possibility to extend the
theory which is conformal in $4D$ to the dimension $D \neq 4$.
It was shown in \cite{anomaly-2004} that the possibility to make
this extension, in the vacuum (i.e., metric-dependent) sector of the
theory, in different ways produces an ambiguity in the local term
in the anomaly-induced action. The purpose of the present work is
to indicate one more ambiguity related to dimensional regularization,
this time concerning the nonlocal action.

There are two ways to apply dimensional regularization. One can
either use it only in the loop integrals, or try to formulate the
theory in the dimension $D \neq 4$ from the very beginning. The last
approach may be more problematic, especially in the case of a gauge
vector field in curved spacetime. The point is that, in $D \neq 4$,
a direct extension of the action does not possess local conformal
invariance. The importance of this problem was noted a long ago,
but the solution which was found in \cite{DeserNepomechie} for
the Abelian gauge field is not very well suited for the quantum
applications because the theory loses gauge invariance. It is
well-known that the simultaneous existence of transverse and
longitudinal modes of a vector field may produce ghost degrees
of freedom in a theory with mass and, probably, in the case of
external gravitational field. More recently, there were found other
ways of extending the vector theory to $D \neq 4$, which guarantee
local conformal symmetry and, at the same time, preserve the gauge
invariance \cite{Asorey21}. It was shown that two out of three
models are closely related. One of these models is very simple and
the conformal symmetry is maintained by introducing a special
auxiliary scalar, called to compensate the effect of changing the
dimension.

In the present work, we explore the conformal anomaly in the
mentioned model with an auxiliary scalar field. One may expect
that, after subtracting divergences and taking the $4D$ limit, this
field should completely disappear. We will show that this does
not happen, such that the remnant of this field remains propagating
in the theory with quantum corrections to the vacuum effective
action.
There are previous works in the literature, with the
calculations of one-loop divergences in a similar models
\cite{Odintsov,Osborn2003}. Indeed, we learned about these
papers only after the main part of the present work was completed.
The main difference of our work is that the scalar field (or
coordinate-dependent couplings), in our case, results from the
formulation of a vector field as conformal theory in the spacetime
dimension $n \neq 4$, as described in \cite{Asorey21}. On the
other hand, technically the one-loop calculations are very similar
and mainly differ by the choice of gauge fixing conditions. We
perform a detailed comparison of all three calculations in the
Appendix C.

The paper is organized as follows. In Sec.~\ref{sec 2} we write down
the action of \cite{Asorey21} in an appropriate way and prepare the
technical background for deriving the one-loop divergences in
Sec.~\ref{sec 3}. Sec.~\ref{sec 4} reports on the calculation of
anomaly and anomaly-induced effective action. Finally, in
Sec.~\ref{sec 5},  we draw our conclusions.
 The main text contains the calculation and its essential steps, while some intermediate expressions are collected in Appendix A. Furthermore, comparisons between the results obtained in this work and those available in the literature for similar models are presented in Appendices B and C.

\section{The conformal vector field model and its bilinear form}
\label{sec 2}

In \cite{Asorey21} four alternative approaches are formulated to
construct a conformal Abelian vector field action. One of the models
is equivalent to the renowned construction of \cite{DeserNepomechie},
which violates the gauge symmetry. In other versions, this symmetry
is preserved. In particular, the non-analytic model
\beq
S^{\star}_d(A,g) \,=\,
-\,\,\frac14\!\int\! d^Dx\sqrt{-g}\,
\big(F_{\mu\nu}F^{\mu\nu}\big)^{D/4},
\label{vec-n-nonloc}
\eeq
is similar to what was proposed earlier in \cite{anomaly-2004} for
the conformal action constructed with the Weyl tensor.
Another model, which is closely related to (\ref{vec-n-nonloc}),
has the action
\beq
S\,=\,-\,\dfrac{1}{4}\int d^Dx\sqrt{-g}\,\psi\,F^{2}_{\mu\nu}\,,
\label{act}
\eeq
where $\psi$ is an auxiliary scalar field, which can also be  regarded
as a coordinate-dependent coupling constant \cite{Osborn2003}. It
is worth noting that an equivalent model was formulated much earlier
in \cite{ETG76}.

In both cases, the electromagnetic field strength tensor is defined
as $F_{\mu\nu}=\pa_{\mu}A_{\nu} - \pa_{\nu}A_{\mu}$, with
$F^{2}_{\mu\nu}=F_{\mu\nu}F^{\mu\nu}$.
In addition to the usual gauge invariance, the action (\ref{act})
is invariant under the local conformal transformation,
\beq
g_{\mu\nu}=\bar{g}_{\mu\nu}e^{2\sigma} \,,
\quad
A_{\mu}=\bar{A}_{\mu} \,,
\quad
\psi=\bar{\psi}\,e^{(4-D)\sigma}
\quad
\textrm{and}
\quad
\sigma=\sigma(x)\,.
\label{eqTC}
\eeq
The transformation rule for $\psi$ compensates the dimension
$D \neq 4$ and the corresponding transformation for $\sqrt{-g}$.
In four dimensions, the auxiliary scalar field does not transform,
leaving the $4D$ action in its conventional form. The conformal
invariance of the action (\ref{act}) make it possible to derive the
contributions of the gauge vector field to the anomaly not only in
$4D$, but also in other dimensions. In particular, two of the
present authors recently reported  on the calculation in $2D$ in
Ref.~\cite{Samuel25}, with some unexpected output, e.g., full
set of the possible terms in the $2D$ anomaly, according to the
classification of \cite{ddi,DeserSchwimmer}, which could be
observed only in four dimensions and beyond, before this
work. Here we explore the same problem in $4D$ and, as can
be seen below, also report on unexpected results.

Using the action (\ref{act}) to derive the anomaly for a quantum
field $A_\mu$ requires, at the first place, calculating the one-loop
divergences around $D=4$. In what follows, we solve this problem
using the heat-kernel method \cite{DeWitt65,DeWitt03}, which is a standard
instrument for quantum field theory in curved space-time.
The Schwinger-DeWitt technique requires the bilinear form of
the action with respect to quantum fields.
Since the action (\ref{act}) is bilinear in $A_{\mu}$, while $\psi$
is treated as a classical background field,  one can skip the use of
the background field method and directly consider quantum $A_{\mu}$.

To provide an efficient removing of the degeneracy, the
Faddeev--Popov procedure requires introducing the
gauge-fixing action depending on the auxiliary field $\psi$,
\beq
S_{\textrm{gf}} \,=\, -\,\dfrac{1}{2}
\int d^{D}x\sqrt{-g}\,\psi\big(\na_{\mu}A^{\mu}\big)^{2}\, .
\label{gf}
\eeq
After some algebra (see \cite{Samuel25} for most of the details),
the bilinear part of the action (\ref{act}), together with the
gauge-fixing term (\ref{gf}), takes the form
\beq
\left(S+S_{\textrm{gf}}\right)^{(2)}
\,=\,
\dfrac{1}{2}\int d^{D}x\sqrt{-g}A^{\mu}H_{\mu}{}^{\nu}A_{\nu}\,,
\label{S2}
\eeq
where the Hessian operator is
\beq
\hat{H} = H_\mu{}^\nu
=\de^{\nu}_{\mu}\cx
-R_{\mu}{}^{\nu}
+\psi^{-1}\Big[
\de^{\nu}_{\mu}(\na^{\la}\psi)\na_{\la}
-(\na^{\nu}\psi)\na_{\mu}
+(\na_{\mu}\psi)\na^{\nu}
\Big]
\, .
\label{eq:2.4}
\eeq
By comparing this expression with the canonical form of the differential
operator employed in the heat-kernel technique (see, e.g., \cite{OUP}
for an introduction),
\beq
\hat{H}\,=\, \hat{1}\cx + 2\hat{h}^{\la}\na_{\la} +\hat{\Pi}\, ,
\label{hessian}
\eeq
we can identify the corresponding components as
\beq
\hat{\Pi}=-R_{\mu}{}^{\nu}
\,\,\quad
\textrm{and}
\qquad
\hat{h}^{\la}
=\dfrac{1}{2}\psi^{-1}
\left[\de^{\nu}_{\mu} \na^{\la}\psi
- \de^{\la}_{\mu} \na^{\nu}\psi
+g^{\nu\la} \na_{\mu}\psi \right]\, .
\label{eq:2.6}
\eeq
At this point, it becomes clear that a more useful parametrization of
the auxiliary scalar is $\psi=\exp \ph$, which will be used below.
The transformation rule for the new scalar $\ph$ can be easily
obtained from (\ref{eqTC}).

\section{One-loop divergences in four dimensions}
\label{sec 3}

The one-loop divergences of the operator (\ref{hessian}) can
be extracted from the expression
\beq
\bar{\Gamma}^{(1)}
\,=\,
\dfrac{i}{2}  \Tr \ln  \hat{H} \,-\,
i \Tr \ln  \hat{H}_{\rm ghost} \,.
\label{eq:3.1}
\eeq
Let us note that the ghost operator $ \hat{H}_{\rm ghost}$ has
the standard form and its contribution does not depend on the
auxiliary scalar field $\ph$. For this reason, we avoid discussing
it in what follows and only take it into account in the final result.

Using the Schwinger proper-time parametrization,
we arrive at the expansion
\beq
\bar{\Gamma}^{(1)}=
-\dfrac{i}{2}\int^{\infty}_{0} \dfrac{ds}{s}
\dfrac{ie^{-\epsilon s}}{\left(4\pi s\right)^{D/2}}
\int d^Dx\sqrt{-g}
\,\tr \sum^{\infty}_{n=0}\left(is\right)^{n}\hat{a}_{n}\left(x\right)\,,
\label{SchDW}
\eeq
where $\hat{a}_{n}$ are the coincidence limits of the
Schwinger-DeWitt coefficients.
Using the dimensional regularization,  the relevant one-loop
divergences
can be written in the form \cite{DeWitt65,DeWitt03,bavi85}
\beq
&&
\bar{\Gamma}^{(1)}_{\textrm{div}}
\,=\,
-\,\dfrac{\mu^{D-4}}{\ep}\int d^D x\sqrt{-g}\tr a_{2} \,,
\label{Gadiv}
\\
&&
\hat{a}_{2}
\,=\, \dfrac{\hat{1}}{180}
\left(R^{2}_{\mu\nu\al\be}
-R^{2}_{\mu\nu}
+\cx R\right)
+\dfrac{1}{2}\hat{P}^{2}
+\dfrac{1}{6}\big(\cx\hat{P}\big)
+\dfrac{1}{12}\hat{S}^{2}_{\mu\nu} \,.
\label{a2}
\eeq
Here $\ep\equiv\left(4\pi\right)^{2}\left(D-4\right)$ is the
parameter of dimensional regularization and $\mu$ is the
renormalization parameter. Direct calculations give
\beq
&&
\hat{P}
\,=\,
\left[ P \right]_{\mu}{}^{\nu}
\,=\, \hat{\Pi} + \dfrac{\hat{1}}{6}R
- \na_{\alpha}\hat{h}^{\alpha} - \hat{h}_{\al}\hat{h}^{\al}\,
\nn
\\
&&
\qquad
= \, -\, R_{\mu}{}^{\nu}
+\dfrac{1}{6}\de^{\nu}_{\mu}R
-\dfrac{1}{2}\de^{\nu}_{\mu}\cx\ph
+\dfrac{\left(D-2\right)}{4}\left(\na_{\mu}\ph\right)\na^{\nu}\ph \, ,
\label{eq: P}
\\
&&
\hat{S}_{\al\be}
\,=\,
\left[S_{\al\be}\right]_{\mu}{}^{\nu}
\,=\,
\mathcal{\hat{R}}_{\al\be}
+\na_{\be}\hat{h}_{\al}
-\na_{\al}\hat{h}_{\be}
+\hat{h}_{\be}\hat{h}_{\al}
-\hat{h}_{\al}\hat{h}_{\be}
\nn
\\
&&
\qquad \quad
=\,\,	- \, R_{\al \be \mu\,\cdot}^{\quad\,\,\,\nu}
	- \dfrac{1}{4}\,\de^{\nu}_{[\be}\, \left(\na_{\al]}\ph\right) \, \na_{\mu}\ph
	- \dfrac{1}{2}\,\de^{\nu}_{[\be}\, \na_{\al]}\na_{\mu}\ph
	+ \dfrac{1}{4}\, g_{\mu[\be}\left(\na_{\al]}\ph\right)  \na^{\nu}\ph
\nn \\
&&
\qquad \qquad \quad
+ \, \dfrac{1}{2}\, g_{\mu[\be} \na_{\al]}\na^{\nu}\ph
+ \dfrac{1}{4}\de^{\nu}_{[\be} g_{\al]\mu}
\left(\na \ph\right)^2\, .
\label{eq: S}
\eeq

Replacing (\ref{eq: S}) and (\ref{eq: P}) in (\ref{Gadiv}), adding
the contribution from the ghost action, we arrive at the one-loop
divergences. For the interested reader, the intermediate formulas
are collected in the appendix A. Taking $D=4$ in the integrand,
we get
\footnote{As it was already mentioned in
the Introduction, the comparison with the previous calculations
\cite{Osborn2003,Odintsov} can be found in the Appendix B and C.
These calculations were performed for more general
models, see also \cite{BuchPletnev}.}
\beq
&&
\bar{\Gamma}{}^{^{(1)}}_{\textrm{div}}
\,=\,
- \,\dfrac{1}{\epsilon}\int d^4x \sqrt{-g}
\left\{ \dfrac{1}{10}C^{2}_{\mu\nu\al\be}
-\dfrac{31}{180}E_{4}
-\dfrac{1}{10}\cx R
-\dfrac{1}{3}\cx^{2}\ph
\right.
\nn \\
 &&
 \qquad \quad
-\,\dfrac{1}{3}R^{\mu\nu}\na_{\mu}\na_{\nu}\ph
 +\dfrac{1}{6}R\cx\ph
 -\dfrac{2}{3}R^{\mu\nu}
 \left(\na_{\mu}\ph\right)\left(\na_{\nu}\ph\right)
 +\dfrac{1}{6}R\left(\na \ph\right)^{2}
 \nn \\
 &&
 \qquad \quad
 +\,\dfrac{5}{12}\left(\cx\ph\right)^{2}
 +\dfrac{1}{6}\left(\na^{\mu}\ph\right)\cx\na_{\mu}\ph
 -\dfrac{1}{12}\left(\na \ph\right)^{2}\cx\ph
 \nn \\
 &&
 \qquad \quad
 \left.
 -\,\dfrac{1}{6}\left(\na^{\mu}\ph\right)
 \left(\na^{\nu}\ph\right)\na_{\mu}\na_{\nu}\ph
 +\dfrac{1}{16}\left(\na \ph\right)^{4}
 \right\} \,.
\label{Gaoneloop}
\eeq
where $(\na \ph)^2 = g^{\mu\nu}\left(\pa_\mu\ph\right)\pa_\nu\ph$.
In the special case $\psi=1$, and consequently $\ph=0$, the previous
expression boils down to the usual one-loop divergences for a
massless vector field (see, e.g.,\cite{OUP}). However, we
are not obliged to assume this value for the auxiliary scalar, and
therefore the dynamic terms with $\ph$ become relevant divergences.

There is a general statement about the conformal invariance of
one-loop divergences in the classically conformal theories  (see,
e.g., \cite{OUP}). This property holds even regardless the conformal
symmetry of the gauge models gets broken under the Faddeev-Popov
procedure \cite{tmf}. Therefore, our model is covered by this
statement. On the other hand, the
form of divergences (\ref{Gaoneloop}) apparently contradicts this
expectation. However, after several integrations by parts,
commutations of covariant derivatives, and using the third Bianchi
identity, one can rewrite the one-loop divergences in the new form
\beq
&&
\bar{\Gamma}{}^{^{(1)}}_{\textrm{div}}
\,=\,
- \,\dfrac{1}{\epsilon}\int d^4x \sqrt{-g}
\,\,\bigg\{
\dfrac{1}{10}C^{2}_{\mu\nu\al\be}
+ \dfrac{1}{4}\ph\De_{4}\ph
+ \dfrac{1}{16}\left(\na \ph\right)^4
\nn \\
 &&
 \qquad \quad
-\dfrac{31}{180}E_{4}
-\dfrac{1}{10}\cx R
-\dfrac{1}{3}\cx^{2}\ph
  -\dfrac{1}{3}\na_{\mu}\left[R^{\mu\nu}\na_{\nu}\ph\right]
 +\dfrac{1}{6}\na_{\mu}\left[R\na^{\mu}\ph\right]
\nn \\
 &&
 \qquad \quad
 +\dfrac{5}{12}\na_{\mu}\left[\left(\na^{\mu}\ph\right)\cx\ph\right]
 -\dfrac{1}{4}\na_{\mu}\left[\ph\na^{\mu}\cx\ph\right]
 -\dfrac{1}{2}\na_{\mu}\left[\ph R^{\mu\nu}\na_{\nu}\ph\right]
 \nn \\
 &&
 \qquad \quad
 +\dfrac{1}{6}\na_{\mu}\left[\ph R\na^{\mu}\ph\right]
 -\dfrac{1}{12}\na_{\nu}\left[\left(\na \ph\right)^2\na^{\nu}\ph\right]
\bigg\} \,.
\label{Gaconf}
\eeq
The first three terms in the integrand are legitimate conformal
($C$-type) invariants and the rest of this expression are $N$-terms,
that means the sum of
the topological Gauss-Bonnet term $E_4$ with a number of
total derivative terms. These terms are not relevant to the dynamics
of the classical theory, but their renormalization may give finite
contributions to the anomaly and the anomaly-induced effective action.
Let us note that these surface terms are not covered by the theorem
of \cite{tmf} and, therefore, our results are consistent with the
expectations.
Another interesting feature of expression (\ref{Gaconf}) is that
the second term in the integrand is based on the well-known Paneitz
operator \cite{Paneitz},
\beq
\De_{4}
\,=\,
\cx^{2}
+\,2R^{\mu\nu}\na_{\mu}\na_{\nu}
\,-\,\dfrac{2}{3}R\cx
\,+\,\dfrac{1}{3}\left(\na^{\mu}R\right)\na_{\mu} .
\label{eq: Paneitz}
\eeq
This fourth-order Hermitian conformally invariant operator was
originally introduced in the framework of conformal supergravity
\cite{FrTs-superconf} and plays an important role in $4D$ quantum
conformal theories and their applications. It is remarkable that
(\ref{Gaconf}) represents a new framework in which this operator
naturally emerges.

\section{Trace anomaly and anomaly-induced action}
\label{sec 4}

The derivation of the trace anomaly may be performed by the method
suggested by Duff \cite{duff77}, with certain simplifications
\cite{PoImpo}. This method works in the case of gauge theory and
dimensional regularization because: \ \textit{i)} the one-loop
divergences are given by the sum of $C$-type and $N$-type conformal
terms \cite{tmf} and because;  \ \textit{ii)} the auxiliary scalar field
$\ph = \ln \psi$ transforms, according to the rule (\ref{eqTC}), as
$\ph=\bar{\ph} + \left(4-D\right)\si$, such that the conformal Noether
identity does not depend on $\ph$ after taking the $4D$ limit. These two
points guarantee that the presence of $\ph$ does not modify the general
scheme for anomaly.

The renormalized one-loop effective action has the form
\beq
\Ga_{R}
\,=\,
S_{\textrm{class}}
\,+\,\bar{\Ga}{}^{(1)}
\,+\,\De S \, ,
\label{eq: AR}
\eeq
where $S_{\textrm{class}}$ is the classical action, $\bar{\Ga}{}^{(1)}=\bar{\Ga}{}^{^{(1)}}_{\textrm{div}}
+\bar{\Ga}{}^{^{(1)}}_{\textrm{fin}}$
is the non-renormalized one-loop quantum correction to the classical
action, and $\De S$ is the infinite local counterterm introduced
to cancel the UV divergences. Owing to the conformal integrand of
the divergences \cite{tmf}, only $\De S$ may produce the violation
of conformal invariance of $\Ga_{R}$ after removing the dimensional
regularization, \cite{duff77,duff94} (see also
\cite{PoImpo,OUP} for details), i.e.,
\beq
\langle T \rangle   \,=\,
- \, \frac{1}{\sqrt{-\bar{g}}}\,e^{-4\si}\,
\frac{\de \De S}{\de \si}\bigg|_{\bar{g}_{\mu\nu}\to g_{\mu\nu},
\,\,\bar{\ph}\to \ph, \,\, \si\to 0}\, .
\label{eq: anomlyfor}
\eeq
Taking into account the conformal transformation rules
(\ref{eqTC}) as well as those for $C_{\mu\nu\al\be}$,
$E_{4}$ and $\cx R$, after taking the $4D$ limit, one
arrives at the trace anomaly,
\beq
&&
\left\langle T\right\rangle
\, =\,- \,\Big\{
w C^{2}_{\mu\nu\al\be} + bE_{4} + c\cx R
+ \be_1 \ph\De_{4}\ph
+ \be_2 \left(\na \ph\right)^{4}
+ \na_\mu \chi^\mu \Big\} \,,
\label{T}
\eeq
where
\beq
&&
\chi^\mu
\,= \, 
\ga_1 R^{\mu\nu}\ph_\nu
+ \ga_2 R \ph^\mu
+ \ga_3 \ph^\mu \cx\ph
+ \ga_4  \ph\na^\mu \cx \ph
\nn
\\
&&
\qquad
\quad
+ \, \ga_5 \ph R^{\mu\nu}\ph_\nu
+ \ga_6 \ph R\ph^\mu
+ \ga_7 \left(\na \ph\right)^2 \ph^\mu
 +  \ga_8 \na^\mu \cx\ph \, .
\label{chimu}
\eeq
In the last expression $\ph_\mu = \na_\mu \ph$,
$\ph^\mu = \bar{g}^{\mu\nu}\ph_\nu$,
$\ph_{\mu\nu} = \na_\mu \na_\nu \ph$, etc. We shall use
these notations, including for $\si$. The coefficients are
\beq
&&
w  = - \, c = \dfrac{1}{10 (4\pi)^2}\,,
\quad
b = - \dfrac{31}{180(4\pi)^2} \, ,
\nn
\\
&&
\be_1 = \dfrac{1}{ 4 (4\pi )^2} \, ,
\quad
\be_{2}=\dfrac{1}{16(4\pi )^2}\, ,
\nn
\\
&&
\ga_1 = - \, 2 \ga_2= - \, 2 \ga_6 = \ga_8  = -\dfrac{1}{3(4\pi )^2}\, ,
\quad
\ga_3 =  \dfrac{5}{12(4\pi )^2}  \, ,
\nn
\\
&&
\ga_4 =  \, \frac12 \ga_5= - \dfrac{1}{4(4\pi )^2}\,,
\quad
\ga_7 =  \dfrac{5}{12(4\pi )^2}\,.
\label{betagama}
\eeq
In this list $w$, $b$ and $c$ are the well-known 
beta functions in the metric sector \cite{Brown77}. In the auxiliary 
scalar sector, we meet two new beta functions $\be_1$ and $\be_2$.
These terms are well-defined in the sense that they do not depend, e.g.,
on the gauge fixing. However, the presence of these terms depends on
our decision to start from the conformal action (\ref{act}), i.e.,
introduce the auxiliary scalar providing the conformal symmetry.
Finally, the coefficients $\ga_k=\ga_1,\,...\,\ga_8$ correspond to
the beta functions of the total derivative terms.h

The anomaly-induced effective action is a solution of the equation
\beq
- \dfrac{2}{\sqrt{- g}}\, g_{\mu\nu}\,
\dfrac{\de \Ga_{\textrm{ind}}(g,\,\ph) }{\de  g_{\mu\nu}}
\,\,=\,\,  \langle T \rangle \, .
\label{eq:4.7}
\eeq
The nonlocal part of $\Ga_{\textrm{ind}}$ can be found in a
standard way, using the formula
\beq
\sqrt{-g}\Big(E_{4}-\dfrac{2}{3}\cx R\Big)
\,=\,
\sqrt{-\bar{g}}
\Big(\bar{E}_{4}-\dfrac{2}{3}\bar{\cx}\bar{R}+\bar{\De}_{4}\si\Big)\,,
\label{eq:4.10}
\eeq
where $\De_{4}$ is defined in (\ref{eq: Paneitz}). It proves useful
to introduce special notation for the $C$-terms,
\beq
Y \big(g_{\mu\nu},\ph\big)
\,=\, wC^2_{\mu\nu\al\be}
+ \be_{1} \ph\De_4\ph
+ \be_{2}(\na \ph)^4 \,.
\label{Y}
\eeq
After this, the nonlocal part of the solution can be written
(see, e.g., \cite{Asorey2022}) in terms of the Green function
$(\sqrt{-g}\De_4)_xG(x,y)=\de(x,y)$, as
\beq
&&
\Ga_{\rm ind-nonloc}
\,=\,
S_c[g_{\mu\nu},\ph]
\,+\,
\frac{b}{8}\iint\limits_{x\,y}\;\Big(E_4
-\frac23\square R\Big)_{\hspace{-1mm}x}
G(x,y)\Big(E_4-\frac23\square R\Big)_{\hspace{-1mm}y}
\nn
\\
&&
\qquad \quad
+\,
\frac{1}{4} \iint\limits_{x\,y}\; Y(x)\, G(x,y)
\Big(E_4-\frac23\square R\Big)_{\hspace{-1mm}y}\,,
\label{nonlocal}
\eeq
where $S_{c}$ is an arbitrary conformal invariant functional
and we use the condensed notation $\int_x = \int d^4x\sqrt{-g}$.
The last expression can be rewritten in the symmetric form
\beq
&&
\Gamma_{\rm ind-nonloc}
\,=\,
S_c[g_{\mu\nu},\ph]
- \frac{1}{8b}\iint\limits_{x\,y}\;Y(x)G(x,y)Y(y)
\nn
\\
&&
\qquad
\quad
+\,\frac{b}{8} \iint\limits_{x\,y}\;\Big(E_4-\frac23\square R+\frac{1}{b}Y\Big)_{\hspace{-1mm}x}G(x,y)\Big(E_4-\frac23\square R+\frac{1}{b}Y\Big)_{\hspace{-1mm}y}.
\label{nonlocal2}
\eeq
One can rewrite the last result  in the local form using auxiliary
fields $\chi$ and $\psi$ \cite{a} (see also \cite{MazMott01} for an
alternative form). The result is
\beq
&&
\Gamma_{\rm  ind-nl-aux}
\,=\,
S_c[g_{\mu\nu},\ph]
\,+\, \frac12  \int_{x}\Big\{\chi\Delta_4\chi - \psi\Delta_4\psi
\nn
\\
&&
\qquad\qquad\qquad
+\,\,
\sqrt{-b}\ph \Big(E_4-\frac23\square R+\frac{1}{b}Y\Big)
\,+\, \frac{1}{\sqrt{-b}}\,\psi Y\Big\}.
\label{aux_fields}
\eeq

Let us note that, compared to \cite{a}, we renamed the first scalar
as $\chi$ to avoid a confusion with the scalar $\ph$ described above, but it is not related with the $\chi^\mu$ of \eqref{chimu}.

Considering the solutions associated with the total-derivative
terms in the anomaly, it is easy to see that these terms in (\ref{T})
and (\ref{chimu}) can be divided into three groups which cannot
mix under the integration. Let us consider them separately.

The unique term of the third order in the auxiliary
scalar is $\ga_7 \na_\mu \big[\ph^\mu (\na \ph)^2\big]$. To deal
with this and other terms we need the first-order transformations
\beq
&&
R\,=\,\bar{R}  - 2\si \bar{R} - 6 \bar{\cx}\si\,,
\nn
\\
&&
R_{\mu\nu}\,=\,\bar{R}_{\mu\nu}  - 2\si_{\mu\nu}
- \bar{g}_{\mu\nu}{\bar\cx}\si\,,
\nn
\\
&&
\cx \ph \,=\,\bar{\cx} \ph  - 2\si \bar{\cx} \ph + 2 \si^\mu \ph_\mu\, ,
\nn
\\
&&
\ph_{\mu\nu} \,=\,\bar{\ph}_{\mu\nu}
- \ph_\mu \si_\nu
- \ph_\nu \si_\mu
+ \bar{g}_{\mu\nu} \si^\la \ph_\la\,.
\label{trans}
\eeq
Note that the first terms in the \textit{r.h.s.}'s of the first three
formulas cancel with the transformation of $\sqrt{-g}$ in the local
actions.

For the mentioned $\ga_7$ - term there are two irreducible local
actions, i.e.,
\beq
&&
\mathcal{L}_{71}
\,=\,(\na \ph)^2 \cx \ph
\qquad
\mbox{and}
\qquad
\mathcal{L}_{72}
\,=\,\ph(\cx \ph)^2\,.
\label{L712}
\eeq
One more possible term
$\mathcal{L}_{73} = \ph \ph^\mu (\na_\mu \cx \ph)$ can be easily
reduced to those in (\ref{L712}) integrating by parts. Using the last
transformation rule in (\ref{trans}), one can easily find that
\beq
\Ga_{ind,\,7} \,=\,\frac12 \int_x \mathcal{L}_{71}
\qquad
\mbox{satisfies}
\qquad
- \,\frac{1}{\sqrt{-g}}\,\frac{\de \Ga_{ind,\,7}}{\de \si}
\,=\, \na_\mu \big[\ph^\mu (\na \ph)^2\big],
\label{Ga7}
\eeq
 and this resolves the part related to $\ga_7$.

Considering now the integration of the terms that are linear in the
auxiliary scalar $\ph$, they have coefficients $\ga_1$,
$\ga_2$, and $\ga_8$. The candidate terms to be part of the
corresponding induced action needs to have four derivatives,
not to be surface integrals and are supposed to be local. The
possible Lagrangians satisfying these conditions are
\beq
\mathcal{L}_{11} = R\cx \ph \,,
\qquad
\mathcal{L}_{12}= R^{\mu\nu} \ph_{\mu\nu} \,,
\qquad
\mathcal{L}_{13} = R^2 \ph\,,
\qquad
\mathcal{L}_{14} = R^2_{\mu\nu}\ph .
\label{L1}
\eeq
Note that $\mathcal{L}_{12}$ can be reduced to
$\mathcal{L}_{11}$ through integrating by parts and using the
third Bianchi identity, hence we do not need to consider this term.
Also, the linear combination of  $\mathcal{L}_{13}$,
$\mathcal{L}_{14}$ and $R^2_{\mu\nu\al\be}\ph$ form the
conformal invariant term. Therefore, the variation of the last
structure is not independent and do not need to be considered.
So, we end up with the three candidate terms only.

The corresponding conformal variations, in the linearized
form, are as follows:
\beq
&&
\de \,\sqrt{-g}\mathcal{L}_{11} \,=\,
2\sqrt{-g}\,\big\{ R \ph^\mu \si_\mu - 3 (\cx \ph)(\cx \si) \big\}\,,
\nn
\\
&&
\de \,\sqrt{-g}\mathcal{L}_{13} \,=\,
-\,12\sqrt{-g}\,R \ph \cx \si\,,
\nn
\\
&&
\de \,\sqrt{-g}\mathcal{L}_{14} \,=\,
-\,2 \sqrt{-g}\,\big\{ 2\ph R^{\mu\nu} \si_{\mu\nu}
+ \ph R \cx \si \big\}.
\label{L1variate}
\eeq

Consider a local term, which is a linear combination with
unknown coefficients $\al_{1k}$,
\beq
\mathcal{L}_{(1)}
\,=\, \al_{11} \mathcal{L}_{11}
+ \al_{13} \mathcal{L}_{13}
+ \al_{14} \mathcal{L}_{14}\,.
\label{L1sum}
\eeq
Using (\ref{L1variate}) and (\ref{L1sum}), after some partial
integrations, we arrive at the equation
\beq
-\dfrac{1}{\sqrt{-g}}\,\dfrac{\de}{\de \si}\int_x \mathcal{L}_{(1)}
\,=\, \na_\mu \chi^\mu_{\,(1)},
\label{L1varichi}
\eeq
where
\beq
&&
\chi^\mu_{\,(1)}
\,=\,
\big(2\al_{11} + 12\al_{13} + 2\al_{14}\big) R \ph^\mu
+ 6\al_{11} \na^\mu \cx  \ph
\nn
\\
&&
\qquad
\quad
+ \,\,4\big(3\al_{13} + \al_{14} \big) \ph \na^\mu R
+ 4 \al_{14} R^{\mu\nu} \ph_\nu
\nn
\\
&&
\qquad
\quad
= \, - \,\ga_1 R^{\mu\nu} \ph_\nu
- \ga_2 R \ph^\mu
- \ga_8   \na^\mu \cx  \ph\,.
\label{chi1}
\eeq
Satisfying this condition is nontrivial because the number of free
parameters is smaller than the number of equations. One can easily
check that the system is compatible only if
$6 \ga_2+ 3\ga_1  = 2\ga_8$.
This condition is not satisfied for the coefficients (\ref{betagama}).
This means, there are no local terms producing the given type of
anomaly terms, in this case.

Finally, consider the most complicated terms which are quadratic
in the scalar $\ph$. In this case, the list of irreducible terms is
\beq
&&
\mathcal{L}_{21} = R (\na \ph)^2 \,,
\qquad
\mathcal{L}_{22}= R^{\mu\nu} \ph_\mu \ph_\nu \,,
\qquad
\mathcal{L}_{23} = R^2 \ph^2\,,
\qquad
\mathcal{L}_{24} = R^2_{\mu\nu}\ph^2 \, ,
\nn
\\
&&
\mathcal{L}_{25} =  (\cx \ph)^2 \,,
\qquad
\mathcal{L}_{26}= R \ph \cx \ph\,,
\qquad
\mathcal{L}_{27} = \ph R^{\mu\nu}\ph_{\mu\nu}\,.
\label{L2}
\eeq
Other possible terms, such as
$ \ph \ph_\mu \na^\mu R$ and $R^2_{\mu\nu\al\be}\ph^2$ are
reducible by means of partial integration or owing to the conformal
symmetry. The transformation rules for the candidate terms are as
follows:
\beq
&&
\de \,\sqrt{-g}\mathcal{L}_{21} \,=\,
- \,6 \sqrt{-g}\, (\na \ph)^2 \cx \si\,,
\nn
\\
&&
\de \,\sqrt{-g}\mathcal{L}_{22} \,=\,
-\,\sqrt{-g}\,\big\{ 2 \ph^\mu \ph^\nu \si_{\mu\nu}
\,+\,
(\na \ph)^2 \cx \si  \big\}\,,
\nn
\\
&&
\de \,\sqrt{-g}\mathcal{L}_{23} \,=\,
-\,12 \sqrt{-g}\,\ph^2 R \cx \si\,,
\nn
\\
&&
\de \,\sqrt{-g}\mathcal{L}_{24} \,=\,
-\,2\sqrt{-g}\,\big\{ 2 \ph^2  R^{\mu\nu} \si_{\mu\nu}
\,+\, \ph^2 R \cx \si  \big\}\,,
\nn
\\
&&
\de \,\sqrt{-g}\mathcal{L}_{25} \,=\,
4\sqrt{-g}\, (\cx \ph) \ph^\mu \si_\mu\,,
\nn
\\
&&
\de \,\sqrt{-g}\mathcal{L}_{26} \,=\,
2\sqrt{-g}\,\big\{ R \ph \ph^\mu \si_\mu
- 3\ph (\cx \ph)(\cx \si) \big\}\,,
\nn
\\
&&
\de \,\sqrt{-g}\mathcal{L}_{27} \,=\,
\sqrt{-g}\,\ph \big\{
R \ph^\mu \si_\mu
- 2 R^{\mu\nu} \ph_\mu \si_\nu
-  (\cx \ph)(\cx \si)
- 2 \ph^{\mu\nu} \si_{\mu\nu}
\big\}\,.
\label{L2variate}
\eeq

As in the previous case, we construct a linear combination
\beq
\mathcal{L}_{(2)}
\,=\, \sum_{k=1}^{7} \al_{2k} \mathcal{L}_{2k}\,.
\label{L2sum}
\eeq
Taking the variational derivative
\beq
-\dfrac{1}{\sqrt{-g}}\,\frac{\de}{\de \si}\int_x \mathcal{L}_{(2)}
\,=\, \na_\mu \chi^\mu_{\,(2)},
\label{L2varichi}
\eeq
we arrive at the equation
\beq
&&
-\,\chi^\mu_{\,(2)}
\,=\,
2\big(6\al_{21} + 2\al_{22} + \al_{27}\big) \ph^{\mu\nu} \ph_\nu
+ \big(2\al_{22}+4\al_{25}+6 \al_{26}+\al_{27}\big) \ph^\mu\cx\ph
\nn
\\
&&
\qquad
\quad
+ \,\big(24\al_{23}+4\al_{24}+ 2\al_{26}+\al_{27}\big) \ph \ph^\mu R
\,+ \, 4\big(3\al_{23} + \al_{24} \big) \ph^2 \na^\mu R
\nn
\\
&&
\qquad
\quad
+ \,8 \al_{24} R^{\mu\nu} \ph_\nu \ph
\,+ \, 3\big(2\al_{26} + \al_{27} \big) \ph \na^\mu \cx \ph
\nn
\\
&&
\qquad
= \, - \,\ga_3 \ph^\mu \cx \ph
- \ga_4 \ph \na^\mu \cx \ph
- \ga_5 \ph R^{\mu\nu} \ph_\nu
- \ga_6 \ph R \ph^\mu\,.
\label{chi2}
\eeq
It turns out that the system of equations for $\al_{2k}$ is
degenerate. The existence of a solution depends on the
identity $6\ga_6 + 3\ga_5 = 2\ga_4$, which is satisfied for
the coefficients (\ref{betagama}). On the other hand, there
is an ambiguity since there are only two equations for
$\al_{21}$, $\al_{22}$, $\al_{25}$, and only one equation
for $\al_{26}$, $\al_{27}$. This means, the local solution
for $\mathcal{L}_{(2)}$ exists, but the coefficients are
ambiguous.

\section{Conclusions}
\label{sec 5}

We have analysed, in the framework of dimensional regularization, 
the one-loop divergences and trace anomaly in the Abelian gauge 
theory formulated as conformal out of $4D$ dimensions. It is known 
for a long time that the vector theory demonstrates certain 
differences compared to the scalar and fermion models. At the first 
place, using the dimensional regularization in a direct way one 
violates the conformal symmetry because for the vector model  $4D$ 
is a very special dimension. According to the existing theorem 
\cite{tmf}, the manner of how dimensional regularization is used 
cannot lead to the nonconformal one-loop divergences. However, the 
anomaly may be affected of how the theory is formulated 
in $D\neq 4$. We have found that, using one of the equivalent ways
of symmetries-preserving extensions \cite{Asorey21}, there are
significant differences with the traditional approaches. The
preservation of both gauge and conformal symmetries requires
either giving up the locality of the theory or introducing an
auxiliary scalar field. In the last case, contrary to the
original expectations, this auxiliary scalar ``survives'' the limit
$D \to 4$ in the finite part of the effective action of vacuum. As
a result, the anomaly-induced effective action in this theory has
not two, but three auxiliary scalars. All three scalars are related
to the Paneitz operator and the new one $\ph$ has the most
complicated interaction term, as can be seen in (\ref{Y}).

The new scalar has one extra surprising feature. For a long time,
the community working on anomalies had an interest to find a
proof of that any total derivative term in the anomaly can be
generated by local Lagrangians in the anomaly-induced effective
action. Usually the discussions of this issue were concentrated
on $6D$ (see, e.g., \cite{Bastianelli:2000}, \cite{InT6d}, and
\cite{Bastianelli:2023}).\footnote{In $4D$ the existence of such
local actions was conformed for all types of the extra background
fields, such as usual scalar or torsion \cite{Asorey2022,AtA}.}
The situation described in the previous section shows that this
proof probably does not exist since we found an explicit 
counterexample of the opposite. 
An important aspect of this finding is that the presence of the 
total derivative terms in the anomaly which cannot be obtained 
from the local lagrangian, is the gauge-fixing dependent 
phenomenon. One can see, in the Appendices B and C, that another 
choice of the gauge-fixing term does not produce such terms, 
in agreement with the previous results \cite{Odintsov,Osborn2003}.

In conclusion, the anomaly in the vector case and dimensional
regularization produces a new surprising ambiguity related to the
way of how the theory is extended to $D \neq 4$. On top of
that, in this case, there is a surprising violation of the common
belief that the total derivative terms in anomaly should be always
produced by local vacuum action.

\section*{Acknowledgements}

Authors are grateful to F. Bastianelli for the discussion of total
derivative terms in the anomaly and to I. Buchbinder for useful
correspondence. S.W.P.O. is grateful to Coordena\c{c}\~{a}o
de Aperfei\c{c}oamento de Pessoal de N\'{i}vel Superior --
CAPES (Brazil) for supporting his current Ph.D. project. The
work of I. Sh. is partially supported by Conselho Nacional de
Desenvolvimento Científico e Tecnológico -- CNPq under the
grant 305122/2023-1.

\section*{Appendix A.
Intermediate expressions for one-loop divergences}
\label{app: A}

Using operators (\ref{eq: S}), one obtains the following
particular traces:
\beq
&&
\tr \cx \hat{P}
\,=\,
- \left(1-\dfrac{D}{6}\right) \cx R
- \left(1-\dfrac{D}{2}\right)\big[
	\left(\na^{\mu}\ph\right)  \cx\na_{\mu}\ph
	+ \left(\na_{\mu}\na_{\nu}\ph\right)^2
	\big]
- \dfrac{D}{2} \cx^2\ph \, ,
\nn
\\
&&
\tr\hat{P}^{2}
\,= \,
R_{\mu\nu}^2
- \dfrac{1}{3}\left(1-\dfrac{D}{12}\right) R^2
+ \dfrac{D}{4}  (\cx\ph)^2
+ \left(1-\dfrac{D}{6}\right)R \cx\ph
+ \dfrac{1}{16} \left(D-2\right)^2 \left(\na \ph\right)^4
\nn
\\
&&
\qquad \qquad
+ \Big(1-\dfrac{D}{2}\Big)
\bigg[
R_{\mu\nu} \,
\left(\na^{\mu}\ph\right)  \na^{\nu}\ph
+ \dfrac{1}{2} 
\left( \na_{\mu}\ph\right)^2  \cx\ph
- \dfrac{1}{6} 
R \left(\na_{\mu}\phi\right)^2\bigg]\,,
\nn
\\
&&
\tr\hat{S}{}^{2}_{\alpha\beta} =
  \, -R_{\mu\nu\al\be}^2
+ R  \left(\na \ph \right)^2
- 4 R_{\mu\nu} \, \na^\mu \na^\nu \ph
- (\cx \phi)^2
- 2 R_{\mu\nu} \left( \na^\mu \ph\right) \na^\nu \ph
 \nn \\
&&
\qquad \qquad
+ \left(2-D\right) \big[
\left(\na_\mu \ph \right)
\left(\na_\nu \ph \right)  \na^\mu \na^\nu \ph
+ \dfrac{1}{8} \left(D-1\right) \left(\na \ph \right)^4
\nn
\\
&&
\qquad \qquad
- \left(\na \ph\right)^2\cx \ph
+  \left( \na_\mu \na_\nu \ph \right)^2 \big]\, .
\eeq
These expressions combine in (\ref{Gaoneloop}) after taking the
$4D$ limit in the integrand.

\section*{Appendix B.
	Ambiguities of the conformal anomaly}
\label{app: B}

The one-loop divergences and conformal anomaly for the action 
(\ref{act}) were previously calculated in \cite{Osborn2003,Odintsov} 
using different approaches. In \cite{Odintsov}, the heat-kernel 
method is combined with the Fujikawa technique to derive the 
conformal anomaly, whereas in \cite{Osborn2003}, the heat-kernel 
method is used together with zeta-function regularization. The most 
important is that the two papers employ different choices of 
gauge-fixing action.

Let us first discuss the differences and similarities between our 
results and those obtained in \cite{Osborn2003}. For the sake of 
comparison, we express the corresponding divergences in the 
same form as in  (\ref{Gaconf}), namely, as a sum of conformal 
invariants and total derivative terms. It turns out that the 
conformal and Gauss-Bonnet terms coincide, so the differences 
concern only the surface terms. 

We shall describe the differences using the coefficients $\ga_k$ 
defined in (\ref{chimu}). Then the main difference is the absence 
of the cubic term proportional to $\ga_{7}$ and the linear term 
with $\ga_{8}$. The remaining linear terms, corresponding to 
$\ga_{1}$ and $\ga_{2}$, are present, 
and the same with the quadratic in $\ph$ terms, associated with 
$\ga_{3}$. Their coefficients, however, differ from those found 
in (\ref{betagama}), while an additional total derivative term is 
present, given by
\beq
	\ga_{\textrm{Osb}}
	\na_{\mu}
	\left[
	\left(\na_{\nu}\ph\right)
	\na^{\mu}\na^{\nu}\ph
	\right]\, .
\eeq
Therefore, in accordance with \cite{Osborn2003}, we obtain the 
following relations
\beq
\ga_1 = -2\ga_2 = 1 \, ,
\quad
\ga_{3} = -\dfrac{1}{4} \, ,
\quad
\ga_{7} = \ga_{8} = 0 \, ,
\quad
\ga_{\textrm{Osb}} = \dfrac{1}{3} \,.
\eeq
All remaining coefficients agree. The situation raises the 
question of whether the difference is related to ambiguities 
in the calculation. As usual with the total derivative terms,
they are ambiguous, that is depend on the regularization 
scheme, for instance they can be modified by the addition 
of local finite counterterms.

Ref.~\cite {Odintsov} also discussed an additional subtlety 
concerning the choice of the gauge-fixing. In general, physical 
quantities should be independent of the choice of gauge-fixing 
action. Nevertheless, the explicit form of the one-loop effective 
action and, consequently, the conformal anomaly may contain 
gauge-dependent terms. This issue is especially relevant for the 
present model because, for the gauge-fixing action (\ref{gf}), 
with $\psi=e^{\ph}$, the factor $\psi$ effectively plays the role 
of a gauge parameter. As discussed in \cite{Odintsov}, this 
leads to an additional factor in the Faddeev--Popov--DeWitt 
determinant,
\beq
\De \,=\,
\det \Big[(-g)^{-\frac{1}{4}}e^{-\frac{1}{2}\ph}\Big]
\det\left[\sqrt{-g}e^{\ph}\cx\right] \, ,
\eeq
thereby affecting the Faddeev--Popov procedure.
It is also worth emphasizing that analogous terms, such as the first 
term on the \textit{r.h.s.} of the equation above, may arise from the 
choice of variables used in the path integral. Since the functional 
measure, together with the gauge-fixing prescription, contributes to 
the definition of the quantum theory, different choices may affect 
the form of the local conformal anomaly. Thus, in the present model, 
the choice of gauge-fixing action can affect the coefficients of the 
terms belonging to the ambiguous sector of the anomaly. This can 
be seen explicitly by repeating the procedure described in 
Sec.~\ref{sec 3} with the gauge-fixing action used in 
\cite{Osborn2003},
\beq
	S_{\textrm{gf}}
	=
	-\dfrac{1}{2}
	\int d^{4}x\sqrt{-g}\,
	\dfrac{1}{\psi}
	\left(\na_{\mu}\psi \, A^{\mu}\right)^{2} \, ,
\eeq
instead of (\ref{gf}) in Sec. \ref{sec 3}. Using the heat-kernel 
technique with this choice, we recover the result of 
\cite{Osborn2003}, including the values of the coefficients 
$\ga_k$, as one can see in Appendix C.

As already stated, the one-loop divergence (\ref{Gaoneloop}) agrees 
with \cite{Odintsov}, although we adopt a different classification of 
the terms appearing in the conformal anomaly. In \cite{Odintsov}, the 
anomaly is classified into trivial and non-trivial contributions. The 
non-trivial sector consists from conformal invariants, 
including the terms linear and quadratic in $\ph$. The trivial 
contribution consists of the total derivative terms. 


All in all, the conformal invariant terms are confirmed to be free 
from ambiguities, which means well-defined non-local part of the
anomaly-induced action. 
On the other hand, the coefficients of the total
derivative terms are sensitive to the choice of regularization, 
functional measure, finite local counterterms, and gauge-fixing. 
Upon integrating the conformal anomaly, these ambiguities are 
transferred to the local part of the anomaly-induced effective 
action. 
For a certain choice of gauge fixing, the total derivative terms
cannot be integrated to the local terms in induced action. 

\section*{Appendix C.
One-loop divergences for another gauge-fixing}
\label{app: C}

For the gauge-fixing action adopted in \cite{Osborn2003}, the 
Hessian operator is given by
\beq
	\hat{H} = H_{\mu}{}^{\nu}
	=\de^{\nu}_{\mu}\cx
	-R_{\mu}{}^{\nu}
	+\dfrac{1}{\psi}\Big[
	\de^{\nu}_{\mu}\left(\na_{\la}\psi\right)\na^{\la}
	+\na_{\mu}\na^{\nu}\psi
	-\dfrac{1}{\psi}\left(\na_{\mu}\psi\right)\na^{\nu}\psi
	\Big]\, ,
\eeq
where we identify
\beq
	\hat{h}^{\la}
	= 	\dfrac{1}{2\psi}\de^{\nu}_{\mu}
	\na^{\la}\psi \, ,
	\quad
	\textrm{and}
	\quad
	\hat{\Pi}=-R_{\mu}{}^{\nu}
	+\dfrac{1}{\psi}\Big[
	\na_{\mu}\na^{\nu}\psi
	-\dfrac{1}{\psi}\left(\na_{\mu}\psi\right)\na^{\nu}\psi
	\Big] \, .
\eeq

The ghost part corresponds to 
\beq
	H_{\textrm{ghost}} = \cx
	+ \dfrac{1}{\psi}\left(\na_{\la}\psi\right)\na^{\la}
	\quad
	\textrm{with}
	\quad
	\hat{h}^{\la} = \dfrac{1}{2\psi}\na^{\la}\psi \, .
\eeq
Taking the parametrization $\psi=\exp\ph$ and applying the 
Schwinger-DeWitt technique as in Sec.~\ref{sec 3}, we obtain
\beq
&&	
\hat{P} = -R_{\mu}{}^{\nu}
	+ \dfrac{1}{6}\de^{\nu}_{\mu}R
	- \dfrac{1}{2}\de^{\nu}_{\mu}\cx\ph
	- \dfrac{1}{4}\de^{\nu}_{\mu}\left(\na_{\la}\ph\right)^{2}
	+ \na_{\mu}\na^{\nu}\ph,
\nn
\\
&&
\hat{S}{}_{\al\be} = -R_{\al\be\mu}{}^{\nu} \,.
\eeq
After a small algebra, instead of the expressions from Appendix A, 
we find 
\beq
&&
\tr \cx\hat{P} = - \Big(1-\dfrac{D}{6}\Big)\cx R
	+ \Big(1-\dfrac{D}{2}\Big)\cx^{2}\ph
	- \dfrac{D}{2}\left[
	\left(\na^{\mu}\ph\right)\cx\na_{\mu}\ph
	+ \left(\na_{\mu}\na_{\nu}\ph\right)^{2}
	\right] \, ,
\nn
\\
&&
\tr\hat{P}^{2} =
	R^{2}_{\mu\nu}
	- \dfrac{1}{3}\Big(1-\dfrac{D}{12}\Big)R^{2}
	+ \dfrac{1}{3}\Big(4-\dfrac{D}{2}\Big)R\cx\ph
	+ \dfrac{1}{2}\Big(1-\dfrac{D}{6}\Big)
	  R\left(\na_{\mu}\ph\right)^{2}
\nn
\\
&&
\qquad \quad
- \,2R^{\mu\nu}\na_{\mu}\na_{\nu}\ph
	- \Big(1-\dfrac{D}{4}\Big)\left(\cx\ph\right)^{2}
	- \dfrac{1}{2}\Big(1-\dfrac{D}{2}\Big)
	\left(\na_{\mu}\ph\right)^{2}\cx\ph
\nn
\\
&&
\qquad \quad
	+ \,\Big(1+\dfrac{D}{16}\Big)\left(\na_{\mu}\ph\right)^{4}\,,
\nn
\\
&&
\tr\hat{S}{}^{2}_{\al\be} = -\, R^{2}_{\mu\nu\al\be}\, .
\eeq

For the ghost contribution, we find
\beq
&&
	\hat{P}_{\textrm{gh}} = \dfrac{1}{6}R
	- \dfrac{1}{2}\cx\ph
	- \dfrac{1}{4}\left(\na_{\la}\ph\right)^{2}\, ,
\qquad 
\hat{S}{}_{\al\be,\,\textrm{gh}} = 0 \,;
\nn
\\
&&
\tr\cx\hat{P}_{\textrm{gh}}
	=
	\dfrac{1}{6}\cx R
	- \dfrac{1}{2}\cx^{2}\ph
	- \dfrac{1}{2}\left[
	\left(\na^{\mu}\ph\right)\cx\na_{\mu}\ph
	+ \left(\na_{\mu}\na_{\nu}\ph\right)^{2}
	\right] \, ,
\nn
\\
&&
	\tr\hat{P}^{2}_{\textrm{gh}}  =
	\dfrac{1}{36}R^{2} - \dfrac{1}{6}R\cx\ph
	- \dfrac{1}{12}R\left(\na_{\mu}\ph\right)^{2}
	+ \dfrac{1}{4}\left(\cx\ph\right)^{2}
	+ \dfrac{1}{4}\left(\na_{\mu}\ph\right)^{2}\cx\ph
\nn
\\
&&
\qquad \quad
	+ \dfrac{1}{16}\left(\na_{\mu}\ph\right)^{4} \, ,
	\quad
	\tr\hat{S}{}^{2}_{\al\be,\,\textrm{gh}} = 0 \, .
\eeq
Replacing the equations above in (\ref{Gadiv}) and taking $D=4$
in the integrand, after integration by parts and algebraic 
manipulations, we obtain the one-loop divergences in the form
\beq
&&
	\bar{\Gamma}{}^{(1)}_{\textrm{div}}
	=
	- \dfrac{1}{\ep}\int d^{4}x\sqrt{-g}\left\{
	 \dfrac{1}{10}C^{2}_{\mu\nu\al\be}
	 + \dfrac{1}{4}\ph\Delta_{4}\ph
	 + \dfrac{1}{16}\left(\na\ph\right)^{4}\right.
\nn
\\
&&
\qquad \quad
	- \dfrac{31}{180}E_{4}-\dfrac{1}{10}\cx R
	- \na_{\mu}\left[R^{\mu\nu}\na_{\nu}\ph\right]
	+ \dfrac{1}{2}\na^{\mu}\left[R\na_{\mu}\ph\right]
\nn
\\
&&
\qquad \quad
	- \dfrac{1}{4}\na_{\mu}\left[\left(\na^{\mu}\ph\right)\cx\ph\right]
	+ \dfrac{1}{3}\na_{\mu}\left[\left(\na_{\nu}\ph\right)
\na^{\mu}\na^{\nu}\ph\right]
	-\dfrac{1}{4}\na_{\mu}\left[\ph\na^{\mu}\cx\ph\right]
\nn
\\
&&
\qquad \quad
	- \dfrac{1}{2}\na_{\mu}\left[\ph R^{\mu\nu}\na_{\nu}\ph\right]
	+ \dfrac{1}{6}\na_{\mu}\left[\ph R\na^{\mu}\ph\right] \, .
\eeq
This result is exactly that of \cite{Osborn2003}. In particular, when 
expressed in the form of (\ref{Gaconf}), it yields the values for 
the coefficients $\ga_i$, which we described in Appendix B, 
\beq
&&
	\ga_1 =  -\dfrac{1}{(4\pi )^2}\, ,
\quad
	 \ga_2  = -2 \ga_3 =  -2 \ga_4 = -\ga_5 =  \dfrac{1}{2(4\pi )^2}  \, ,
\nn
\\
&&
	\ga_\textrm{Osb} =  2 \ga_6=  \dfrac{1}{3(4\pi )^2}\,,
\quad
	\ga_7 = \ga_8 = 0 \, .
\eeq


\end{document}